\documentclass[aps,preprint,nofootinbib,superscriptaddress]{revtex4}%
\usepackage{amssymb}
\usepackage{amsmath}
\usepackage{epsfig}
\usepackage{amsfonts}
\usepackage{graphicx}
\usepackage{hyperref}%
\providecommand{\U}[1]{\protect\rule{.1in}{.1in}}
\begin{document}
\title{Recurrence Relations of Higher Spin BPST Vertex Operators for Open String }
\author{Chih-Hao Fu}
\email{zhihaofu@nctu.edu.tw}
\affiliation{Department of Electrophysics, National Chiao-Tung University and Physics
Division, National Center for Theoretical Sciences, Hsinchu, Taiwan, R.O.C.}
\author{Jen-Chi Lee}
\email{jcclee@cc.nctu.edu.tw}
\affiliation{Department of Electrophysics, National Chiao-Tung University and Physics
Division, National Center for Theoretical Sciences, Hsinchu, Taiwan, R.O.C.}
\author{Chung-I Tan}
\email{chung-i\_tan@brown.edu}
\affiliation{Physics Department, Brown University, Providence, RI 02912, USA}
\author{Yi Yang}
\email{yiyang@mail.nctu.edu.tw}
\affiliation{Department of Electrophysics, National Chiao-Tung University and Physics
Division, National Center for Theoretical Sciences, Hsinchu, Taiwan, R.O.C.}
\date{\today }

\begin{abstract}
We calculate higher spin BPST vertex operators for open bosonic string and
express these operstors in terms of Kummer function of the second kind. We
derive infinite number of recurrence relations among BPST vertex operators of
different string states. These recurrence relations among BPST vertex
operators lead to the recurrence relations among Regge string scattering
amplitudes discovered recently.

\end{abstract}
\preprint{Brown-HET-1642}
\maketitle
\tableofcontents
%

\setcounter{equation}{0}
\renewcommand{\theequation}{\arabic{section}.\arabic{equation}}%

\section{\bigskip Introduction\bigskip}

Recently there has been interest to study Regge regime (RR) string scattering
amplitudes \cite{LY,BPST, CCW,RR1,Tan1,Tan2} for higher spin string states
\cite{bosonic,RRsusy,LY,Sagnotti,Moore}. One of the motivation was to
understand their intimate link with the scattering amplitudes in the fixed
angle or Gross regime (GR) \cite{GM,GrossManes,Gross,West1,West2}. In the GR,
a saddle point method was used to calculate string-tree amplitudes
\cite{ChanLee1,ChanLee2,PRL,CHLTY}, and the ratios of scattering amplitudes
among different string states at each fixed mass level can be extracted and
were found to be independent of scattering energy and scattering angle.
Alternatively, these ratios can be rederived algebraically by solving linear
relations or GR stringy Ward identities from decoupling of zero-norm states
(ZNS) \cite{ZNS1,ZNS3,ZNS2}. More interestingly, these infinite number of
ratios for the GR can be extracted from RR string scattering amplitudes based
on summation algorithms for Stirling number identities \cite{LYAM,MK}.

In contrast to the GR, an infinite number of recurrence relations among higher
spin RR string scattering amplitudes were discovered more recently \cite{LY}.
Instead of RR stringy Ward identities derived from decoupling of ZNS, the
calculation was based on recurrence relations of Kummer functions of the
second kind \cite{Slater}. These recurrence relations among RR amplitudes were
considered to be dual to the linear relations among GR amplitudes discussed above.

In this paper, we study higher spin Regge string scattering amplitudes from
BPST vertex operator approach. Note that in the original BPST paper
\cite{BPST}, the authors calculated the case of closed string and thus Pomeron
vertex operators. Here, for simplicity, we will calculate higher spin BPST
vertex operators at arbitrary mass levels of open bosonic string
\footnote{Taking advantage of Regge factorization, a Pomeron vertex operator
$\mathcal{V}_{P}$ was introduced in \cite{BPST}, which allows one to calculate
the coupling between the leading closed string Regge trajectory with any
n-particle external state $|\mathcal{W}\rangle$. In this paper, we only
consider 4-point scattering for open strings. As such, we only need to treat
the coupling of the leading open-string Reggeon to two-particle states. For
brevity, we use here the term \textquotedblleft higher spin BPST vertex
operators" collectively for the product of the vertex operator for the leading
open string Reggeon with external two-particle states, one of which with high
spin.}. The calculation can be easily generalized to closed string case. We
find that all BPST vertex operators can be expressed in terms of Kummer
functions of the second kind. We can then derive infinite number of recurrence
relations among BPST vertex operators of different string states. These
recurrence relations among BPST vertex operators lead to the recurrence
relations among Regge string scattering amplitudes discovered recently
\cite{LY}.

\section{Four Tachyon Scattering}

We will calculate high energy open string scatterings in the Regge Regime
\begin{equation}
s\rightarrow\infty,\sqrt{-t}=\text{fixed (but }\sqrt{-t}\neq\infty)
\end{equation}
where%
\begin{equation}
s=-(k_{1}+k_{2})^{2}\text{ and }t=-(k_{2}+k_{3})^{2}.
\end{equation}
Note that the convention for $s$ and $t$ adopted here is different from the
original BPST paper in \cite{BPST}.

We first review the calculation of tachyon BPST vertex operator \cite{BPST}.
The $s-t$ channel of open string four tachyon amplitude can be written as%
\begin{equation}
A=\int_{0}^{1}d\omega\cdot\omega^{k_{1}\cdot k_{2}}\left(  1-\omega\right)
^{k_{2}\cdot k_{3}}=\int_{0}^{1}d\omega\cdot\omega^{-2-\frac{s}{2}}\left(
1-\omega\right)  ^{-2-\frac{t}{2}}.
\end{equation}
Since $s\rightarrow\infty$, the integral is dominated around $\omega=1$.
Making the variable transformation $\omega=1-x$, the integral is dominated
around $x=0$, we obtain%
\begin{equation}
A=\int_{0}^{1}dx\cdot\left(  1-x\right)  ^{-2-\frac{s}{2}}x^{-2-\frac{t}{2}%
}\simeq\int dx\cdot x^{-2-\frac{t}{2}}e^{\frac{s}{2}x}=\Gamma\left(
-1-\frac{t}{2}\right)  \left(  -\frac{s}{2}\right)  ^{1+\frac{t}{2}}.
\label{Regge}%
\end{equation}
Alternatively, the integral in $A$ can be expressed as%
\begin{equation}
A=\int d\omega\left\langle e^{ik_{1}X(0)}e^{ik_{2}X(\omega)}e^{ik_{3}%
X(1)}e^{ik_{4}X(\infty)}\right\rangle . \label{AMP}%
\end{equation}
One can calculate the operator product expansion (OPE) in the Regge limit%
\begin{equation}
e^{ik_{2}X(w)}e^{ik_{3}X(z)}\sim\left\vert w-z\right\vert ^{k_{2}\cdot k_{3}%
}e^{i(k_{2}+k_{3})X(z)+ik_{2}(w-z)\partial X(z)+\cdots}.\nonumber
\end{equation}
This means%
\begin{equation}
e^{ik_{2}X(\omega)}e^{ik_{3}X(1)}\sim\left(  1-\omega\right)  ^{k_{2}\cdot
k_{3}}e^{ikX(1)-ik_{2}\left(  1-\omega\right)  \partial X(1)+higher\text{
}power\text{ }of\text{ }\left(  1-\omega\right)  },k=k_{2}+k_{3}.
\label{OPE-tachyon}%
\end{equation}
In evaluating Eq.(\ref{AMP}), one can instead carry out the $\omega$
integration first in Eq.(\ref{OPE-tachyon}) at the operator level to obtain
the BPST vertex operator \cite{BPST}%
\begin{align}
V_{BPST}  &  =\int d\omega e^{ik_{2}X(\omega)}e^{ik_{3}X(1)}\nonumber\\
&  \sim\int d\omega\left(  1-\omega\right)  ^{k_{2}\cdot k_{3}}%
e^{ikX(1)-ik_{2}\left(  1-\omega\right)  \partial X(1)}\nonumber\\
&  =\int dxx^{k_{2}\cdot k_{3}}e^{ikX(1)-ik_{2}x\partial X(1)}\nonumber\\
&  =\Gamma\left(  -1-\frac{t}{2}\right)  \left[  ik_{2}\partial X(1)\right]
^{1+\frac{t}{2}}e^{ikX(1)},
\end{align}
which leads to the same amplitude as in Eq.(\ref{Regge})%
\begin{align}
A  &  =\left\langle e^{ik_{1}X(0)}V_{P}e^{ik_{4}X(\infty)}\right\rangle
\nonumber\\
&  =\Gamma\left(  -1-\frac{t}{2}\right)  \left\langle e^{ik_{1}X(0)}\left[
ik_{2}\partial X(1)\right]  ^{1+\frac{t}{2}}e^{ikX(1)}e^{ik_{4}X(\infty
)}\right\rangle \nonumber\\
&  =\Gamma\left(  -1-\frac{t}{2}\right)  \left(  k_{1}k_{2}\right)
^{1+\frac{t}{2}}\nonumber\\
&  \sim\Gamma\left(  -1-\frac{t}{2}\right)  \left(  -\frac{s}{2}\right)
^{1+\frac{t}{2}}.
\end{align}

\section{Higher Spin BPST Vertex}

\subsection{A spin two state}

It was shown \cite{bosonic,RRsusy,LY} that for the 26D open bosonic string
states of leading order in energy in the Regge limit at mass level $M_{2}%
^{2}=2(N-1)$, $N=\sum_{n,m,l>0}np_{n}+mq_{m}+lr_{l}$ are of the form (we
choose the second state of the four-point function to be the higher spin
string state)%
\begin{equation}
\left\vert p_{n},q_{m},r_{l}\right\rangle =\prod_{n>0}(\alpha_{-n}^{T}%
)^{p_{n}}\prod_{m>0}(\alpha_{-m}^{P})^{q_{m}}\prod_{l>0}(\alpha_{-l}%
^{L})^{r_{l}}|0,k\rangle\label{RR}%
\end{equation}
where the polarizations of the 2nd particle with momentum $k_{2}$ on the
scattering plane were defined to be $e^{P}=\frac{1}{M_{2}}(E_{2}%
,\mathrm{k}_{2},0)=\frac{k_{2}}{M_{2}}$ as the momentum polarization,
$e^{L}=\frac{1}{M_{2}}(\mathrm{k}_{2},E_{2},0)$ the longitudinal polarization
and $e^{T}=(0,0,1)$ the transverse polarization which lies on the scattering
plane. $\eta_{\mu\nu}=diag(-1,1,1)$. The three vectors $e^{P}$, $e^{L}$ and
$e^{T}$ satisfy the completeness relation $\eta_{\mu\nu}=\sum_{\alpha,\beta
}e_{\mu}^{\alpha}e_{\nu}^{\beta}\eta_{\alpha\beta}$ where $\mu,\nu=0,1,2$ and
$\alpha,\beta=P,L,T$ and $\alpha_{-1}^{T}=\sum_{\mu}e_{\mu}^{T}\alpha
_{-1}^{\mu}$, $\alpha_{-1}^{T}\alpha_{-2}^{L}=\sum_{\mu,\nu}e_{\mu}^{T}e_{\nu
}^{L}\alpha_{-1}^{\mu}\alpha_{-2}^{\nu}$ etc.

In this section, we first consider a simple case of a spin two state
$\alpha_{-1}^{P}\alpha_{-1}^{P}|0\rangle$ corresponding to the vertex $\left(
\partial X^{P}\right)  ^{2}e^{ik_{2}X}\left(  \omega\right)  $. The four-point
amplitude of the spin two state with three tachyons can be calculated by using
the conventinal method%
\begin{align}
A^{(q_{1}=2)}  &  =\int d\omega\left\langle e^{ik_{1}X(0)}\left(  \partial
X^{P}\right)  ^{2}e^{ik_{2}X}\left(  \omega\right)  e^{ik_{3}X(1)}%
e^{ik_{4}X(\infty)}\right\rangle \nonumber\\
&  =\int d\omega\omega^{k_{1}\cdot k_{2}}(1-\omega)^{k_{2}\cdot k_{3}}\left[
\frac{ie^{P}\cdot k_{1}}{-\omega}+\frac{ie^{P}\cdot k_{3}}{1-\omega}\right]
^{2}\nonumber\\
&  =-(e^{P}\cdot k_{1})^{2}\Gamma\left(  -1-\frac{t}{2}\right)  \left(
-\frac{s}{2}\right)  ^{\frac{t}{2}-1}+2(e^{P}\cdot k_{1})(e^{P}\cdot
k_{3})\Gamma\left(  -2-\frac{t}{2}\right)  \left(  -\frac{s}{2}\right)
^{\frac{t}{2}}\nonumber\\
&  -(e^{P}\cdot k_{3})^{2}\Gamma\left(  -3-\frac{t}{2}\right)  \left(
-\frac{s}{2}\right)  ^{\frac{t}{2}+1}. \label{amplitude}%
\end{align}
The momenta of the four particles on the scattering plane are%

\begin{align}
k_{1}  &  =\left(  +\sqrt{p^{2}+M_{1}^{2}},-p,0\right)  ,\\
k_{2}  &  =\left(  +\sqrt{p^{2}+M_{2}^{2}},+p,0\right)  ,\\
k_{3}  &  =\left(  -\sqrt{q^{2}+M_{3}^{2}},-q\cos\phi,-q\sin\phi\right)  ,\\
k_{4}  &  =\left(  -\sqrt{q^{2}+M_{4}^{2}},+q\cos\phi,+q\sin\phi\right)
\end{align}
where $p\equiv\left\vert \mathrm{\vec{p}}\right\vert $, $q\equiv\left\vert
\mathrm{\vec{q}}\right\vert $ and $k_{i}^{2}=-M_{i}^{2}$. The relevant
kinematics in the Regge limit are \cite{bosonic,RRsusy,LY}%
\begin{equation}
e^{P}\cdot k_{1}\simeq-\frac{s}{2M_{2}},\text{ \ }e^{P}\cdot k_{3}\simeq
-\frac{\tilde{t}}{2M_{2}}=-\frac{t-M_{2}^{2}-M_{3}^{2}}{2M_{2}}; \label{st}%
\end{equation}%
\begin{equation}
e^{L}\cdot k_{1}\simeq-\frac{s}{2M_{2}},\text{ \ }e^{L}\cdot k_{3}\simeq
-\frac{\tilde{t}^{\prime}}{2M_{2}}=-\frac{t+M_{2}^{2}-M_{3}^{2}}{2M_{2}};
\label{st2}%
\end{equation}
and%
\begin{equation}
e^{T}\cdot k_{1}=0\text{, \ \ }e^{T}\cdot k_{3}\simeq-\sqrt{-{t}}%
\end{equation}
where $\tilde{t}$ and $\tilde{t}^{\prime}$ are related to $t$ by finite mass
square terms%
\begin{equation}
\tilde{t}=t-M_{2}^{2}-M_{3}^{2}\text{ , \ }\tilde{t}^{\prime}=t+M_{2}%
^{2}-M_{3}^{2}.
\end{equation}
By using Eq.(\ref{st}), one easily see that the three terms in
Eq.(\ref{amplitude}) share the same order of energy in the Regge limit. We
stress that this key observation on the polarizations for higher spin states
was not discussed in \cite{BPST,CCW}.

One can calculate the OPE in the Regge limit%
\begin{equation}
\partial X^{P}\partial X^{P}e^{ik_{2}X}\left(  w\right)  e^{ik_{3}X}\left(
z\right)  \sim\left\vert w-z\right\vert ^{k_{2}\cdot k_{3}}\left[  \partial
X\left(  z\right)  ^{P}+\frac{ie^{P}\cdot k_{3}}{w-z}\right]  ^{2}%
e^{ikX(z)+ik_{2}\left(  w-z\right)  \partial X(z)}.\nonumber
\end{equation}
This means%
\begin{equation}
\partial X^{P}\partial X^{P}e^{ik_{2}X}\left(  \omega\right)  e^{ik_{3}%
X}\left(  1\right)  \sim\left(  1-\omega\right)  ^{k_{2}\cdot k_{3}}\left[
\partial X\left(  1\right)  ^{P}-\frac{ie^{P}\cdot k_{3}}{1-\omega}\right]
^{2}e^{ikX(1)-ik_{2}\left(  1-\omega\right)  \partial X(1)},k=k_{2}+k_{3}.
\label{OPE-vector}%
\end{equation}
One can carry out the $\omega$ integration in Eq.(\ref{OPE-vector}) at the
operator level to obtain the BPST vertex operator%
\begin{align}
V_{BPST}^{(q_{1}=2)}  &  =\int d\omega(\partial X^{P})^{2}e^{ik_{2}X}\left(
\omega\right)  e^{ik_{3}X}\left(  1\right) \nonumber\\
&  \sim\int d\omega\left(  1-\omega\right)  ^{k_{2}\cdot k_{3}}\left[
\partial X\left(  1\right)  ^{P}-\frac{ie^{P}\cdot k_{3}}{1-\omega}\right]
^{2}e^{ikX(1)-ik_{2}\left(  1-\omega\right)  \partial X(1)}\nonumber\\
&  =\partial X\left(  1\right)  ^{P}\partial X\left(  1\right)  ^{P}\int
dxx^{k_{2}\cdot k_{3}}e^{ikX(1)-ik_{2}x\partial X(1)}\nonumber\\
&  -2ie^{P}\cdot k_{3}\partial X\left(  1\right)  ^{P}\int dxx^{k_{2}\cdot
k_{3}-1}e^{ikX(1)-ik_{2}x\partial X(1)}\nonumber\\
&  -(e^{P}\cdot k_{3})^{2}\int dxx^{k_{2}\cdot k_{3}-2}e^{ikX(1)-ik_{2}%
x\partial X(1)}\nonumber\\
&  =\Gamma\left(  -1-\frac{t}{2}\right)  \left[  ik_{2}\partial X(1)\right]
^{\frac{t}{2}-1}\partial X\left(  1\right)  ^{P}\partial X\left(  1\right)
^{P}e^{ikX(1)}\nonumber\\
&  -2ie^{P}\cdot k_{3}\Gamma\left(  -2-\frac{t}{2}\right)  \left[
ik_{2}\partial X(1)\right]  ^{\frac{t}{2}}\partial X\left(  1\right)
^{P}e^{ikX(1)}\nonumber\\
&  -(e^{P}\cdot k_{3})^{2}\Gamma\left(  -3-\frac{t}{2}\right)  \left[
ik_{2}\partial X(1)\right]  ^{\frac{t}{2}+1}e^{ikX(1)} \label{three}%
\end{align}
which leads to the same amplitude%
\begin{align}
A^{(q_{1}=2)}  &  =\left\langle e^{ik_{1}X(0)}V_{BPST}^{(q_{1}=2)}%
e^{ik_{4}X(\infty)}\right\rangle \nonumber\\
&  =\Gamma\left(  -1-\frac{t}{2}\right)  \left\langle e^{ik_{1}X(0)}\left[
ik_{2}\partial X(1)\right]  ^{\frac{t}{2}-1}\partial X\left(  1\right)
^{P}\partial X\left(  1\right)  ^{P}e^{ikX(1)}e^{ik_{4}X(\infty)}\right\rangle
\nonumber\\
&  -2ie^{P}\cdot k_{3}\Gamma\left(  -2-\frac{t}{2}\right)  \left\langle
e^{ik_{1}X(0)}\left[  ik_{2}\partial X(1)\right]  ^{\frac{t}{2}}\partial
X\left(  1\right)  ^{P}e^{ikX(1)}e^{ik_{4}X(\infty)}\right\rangle \nonumber\\
&  -(e^{P}\cdot k_{3})^{2}\Gamma\left(  -3-\frac{t}{2}\right)  \left\langle
e^{ik_{1}X(0)}\left[  ik_{2}\partial X(1)\right]  ^{\frac{t}{2}+1}%
e^{ikX(1)}e^{ik_{4}X(\infty)}\right\rangle \nonumber\\
&  \sim-(e^{P}\cdot k_{1})^{2}\Gamma\left(  -1-\frac{t}{2}\right)  \left(
-\frac{s}{2}\right)  ^{\frac{t}{2}-1}+2(e^{P}\cdot k_{1})(e^{P}\cdot
k_{3})\Gamma\left(  -2-\frac{t}{2}\right)  \left(  -\frac{s}{2}\right)
^{\frac{t}{2}}\nonumber\\
&  -(e^{P}\cdot k_{3})^{2}\Gamma\left(  -3-\frac{t}{2}\right)  \left(
-\frac{s}{2}\right)  ^{\frac{t}{2}+1}. \label{same}%
\end{align}
Note that the three terms in Eq.(\ref{three}) lead to the three terms
respectively in Eq.(\ref{same}) with the same order of energy in the Regge limit.

\subsection{Higher spin states}

We now consider the higher spin state%
\begin{equation}
\left\vert p_{n},q_{m}\right\rangle =\prod_{n=1}(\alpha_{-n}^{T})^{p_{n}}%
\prod_{m=1}(\alpha_{-m}^{P})^{q_{m}}|0\rangle, \label{state1}%
\end{equation}
which corresponds to the vertex%
\begin{equation}
V_{2}\left(  \omega\right)  =\left[  \prod_{n=1}\left(  \partial^{n}%
X^{T}\right)  ^{p_{n}}\prod_{m=1}\left(  \partial^{m}X^{P}\right)  ^{q_{m}%
}\right]  e^{ik_{2}X}\left(  \omega\right)  .
\end{equation}
The four-point amplitude of the above state with three tachyons was calculated
to be (from now on we set $M_{2}=M$) \cite{bosonic,RRsusy,LY}
\begin{align}
A^{(p_{n},q_{m})}  &  =\int d\omega\left\langle e^{ik_{1}X(0)}V_{2}\left(
\omega\right)  e^{ik_{3}X(1)}e^{ik_{4}X(\infty)}\right\rangle \nonumber\\
&  =\left(  -\frac{1}{M}\right)  ^{q_{1}}U\left(  -q_{1},\frac{t}{2}%
+2-q_{1},\frac{\tilde{t}}{2}\right)  B\left(  -1-\frac{s}{2},-1-\frac{t}%
{2}\right) \nonumber\\
&  \cdot\prod_{n=1}\left[  \sqrt{-t}(n-1)!\right]  ^{p_{n}}\prod_{m=2}\left[
\tilde{t}(m-1)!\left(  -\frac{1}{2M}\right)  \right]  ^{q_{m}}\\
&  \sim\left(  -\frac{1}{M}\right)  ^{q_{1}}U\left(  -q_{1},\frac{t}%
{2}+2-q_{1},\frac{\tilde{t}}{2}\right)  \Gamma\left(  -1-\frac{t}{2}\right)
\left(  -\frac{s}{2}\right)  ^{1+\frac{t}{2}}\label{power}\\
&  \cdot\prod_{n=1}\left[  \sqrt{-t}(n-1)!\right]  ^{p_{n}}\prod_{m=2}\left[
\tilde{t}(m-1)!\left(  -\frac{1}{2M}\right)  \right]  ^{q_{m}}
\label{amplitude-tensor}%
\end{align}
where $U$ is the Kummer function of the second kind and is defined to be%
\begin{equation}
U(a,c,x)=\frac{\pi}{\sin\pi c}\left[  \frac{M(a,c,x)}{(a-c)!(c-1)!}%
-\frac{x^{1-c}M(a+1-c,2-c,x)}{(a-1)!(1-c)!}\right]  ,\text{ \ }(c\neq
2,3,4...). \label{M}%
\end{equation}
In Eq.(\ref{M}) $M(a,c,x)=\sum_{j=0}^{\infty}\frac{(a)_{j}}{(c)_{j}}%
\frac{x^{j}}{j!}$ is the Kummer function of the first kind. Here
$(a)_{j}=a(a+1)(a+2)...(a+j-1)$ is the Pochhammer symbol.

One can calculate the OPE in the Regge limit%
\begin{align}
&  V_{2}\left(  \omega\right)  e^{ik_{3}X(1)}\nonumber\\
&  =\left[  \prod_{n=1}\left(  \partial^{n}X^{T}\right)  ^{p_{n}}\prod
_{m=1}\left(  \partial^{m}X^{P}\right)  ^{q_{m}}\right]  e^{ik_{2}X}\left(
\omega\right)  e^{ik_{3}X(1)}\nonumber\\
&  \sim\prod_{n=1}\left[  \frac{\left(  n-1\right)  !k_{3}\cdot e^{T}}{\left(
1-\omega\right)  ^{n}}\right]  ^{p_{n}}\prod_{m=2}\left[  \frac{\left(
m-1\right)  !k_{3}\cdot e^{P}}{\left(  1-\omega\right)  ^{m}}\right]  ^{q_{m}%
}\nonumber\\
&  \cdot\left[  \partial X\left(  1\right)  \cdot e^{P}-\frac{ik_{3}\cdot
e^{P}}{1-\omega}\right]  ^{q_{1}}\left(  1-\omega\right)  ^{k_{2}\cdot k_{3}%
}e^{ikX(1)-ik_{2}\left(  1-\omega\right)  \partial X(1)}\\
&  =\left(  \frac{-\tilde{t}}{2M}\right)  ^{q_{1}}\prod_{n=1}\left[  \sqrt
{-t}(n-1)!\right]  ^{p_{n}}\prod_{m=2}\left[  \tilde{t}(m-1)!\left(  -\frac
{1}{2M}\right)  \right]  ^{q_{m}}\nonumber\\
&  \cdot\sum_{j=0}^{q_{1}}{\binom{q_{1}}{j}}\left(  \frac{2iM_{2}\partial
X\left(  1\right)  \cdot e^{P}}{\tilde{t}}\right)  ^{j}\left(  1-\omega
\right)  ^{k_{2}\cdot k_{3}-N+j}e^{ikX(1)-ik_{2}\left(  1-\omega\right)
\partial X(1)} \label{OPE-tensor}%
\end{align}
\ where $N=\sum_{n,m}\left(  np_{n}+mq_{m}\right)  $. We can carry out the
$\omega$ integration in Eq.(\ref{OPE-tensor}) to obtain the BPST vertex
operator%
\begin{align}
V_{BPST}^{(p_{n};q_{m})}  &  =\int d\omega V_{2}\left(  \omega\right)
e^{ik_{3}X}\left(  1\right) \nonumber\\
&  \sim\left(  \frac{-\tilde{t}}{2M}\right)  ^{q_{1}}\prod_{n=1}\left[
\sqrt{-t}(n-1)!\right]  ^{p_{n}}\prod_{m=2}\left[  \tilde{t}(m-1)!\left(
-\frac{1}{2M}\right)  \right]  ^{q_{m}}\nonumber\\
&  \cdot\sum_{j=0}^{q_{1}}{\binom{q_{1}}{j}}\left(  \frac{2iM_{2}\partial
X\left(  1\right)  \cdot e^{P}}{\tilde{t}}\right)  ^{j}\int d\omega\left(
1-\omega\right)  ^{k_{2}\cdot k_{3}-N+j}e^{ikX(1)-ik_{2}\left(  1-\omega
\right)  \partial X(1)}\nonumber\\
&  =\left(  \frac{-\tilde{t}}{2M}\right)  ^{q_{1}}\prod_{n=1}\left[  \sqrt
{-t}(n-1)!\right]  ^{p_{n}}\prod_{m=2}\left[  \tilde{t}(m-1)!\left(  -\frac
{1}{2M}\right)  \right]  ^{q_{m}}\nonumber\\
&  \cdot\sum_{j=0}^{q_{1}}{\binom{q_{1}}{j}}\left(  \frac{2iM\partial X\left(
1\right)  \cdot e^{P}}{\tilde{t}}\right)  ^{j}\int dxx^{k_{2}\cdot k_{3}%
-N+j}e^{ikX(1)-ik_{2}x\partial X(1)}\nonumber\\
&  =\left(  \frac{-\tilde{t}}{2M}\right)  ^{q_{1}}\prod_{n=1}\left[  \sqrt
{-t}(n-1)!\right]  ^{p_{n}}\prod_{m=2}\left[  \tilde{t}(m-1)!\left(  -\frac
{1}{2M}\right)  \right]  ^{q_{m}}\nonumber\\
&  \cdot\sum_{j=0}^{q_{1}}{\binom{q_{1}}{j}}\left(  \frac{2iM\partial X\left(
1\right)  \cdot e^{P}}{\tilde{t}}\right)  ^{j}\Gamma\left(  -1-\frac{t}%
{2}+j\right)  \left[  ik_{2}\cdot\partial X(1)\right]  ^{1+\frac{t}{2}%
-j}e^{ikX(1)}. \label{dot}%
\end{align}
One notes that, in Eq.(\ref{dot}), $M\partial X\left(  1\right)  \cdot
e^{P}=k_{2}\cdot\partial X(1)$ and the summation over $j$ can be simplified.
The BPST vertex operator can be further reduced to
\begin{align}
V_{BPST}^{(p_{n};q_{m})}  &  =\left(  \frac{-\tilde{t}}{2M_{2}}\right)
^{q_{1}}\prod_{n=1}\left[  \sqrt{-t}(n-1)!\right]  ^{p_{n}}\prod_{m=2}\left[
\tilde{t}(m-1)!\left(  -\frac{1}{2M}\right)  \right]  ^{q_{m}}\nonumber\\
&  \cdot\sum_{j=0}^{q_{1}}{\binom{q_{1}}{j}}\left(  \frac{2}{\tilde{t}%
}\right)  ^{j}\left(  -1-\frac{t}{2}\right)  _{j}\Gamma\left(  -1-\frac{t}%
{2}\right)  \left[  ik_{2}\cdot\partial X(1)\right]  ^{1+\frac{t}{2}%
}e^{ikX(1)}\nonumber\\
&  =\left(  \frac{-1}{M}\right)  ^{q_{1}}\prod_{n=1}\left[  \sqrt
{-t}(n-1)!\right]  ^{p_{n}}\prod_{m=2}\left[  \tilde{t}(m-1)!\left(  -\frac
{1}{2M}\right)  \right]  ^{q_{m}}\nonumber\\
&  \cdot U\left(  -q_{1},\frac{t}{2}+2-q_{1},\frac{\tilde{t}}{2}\right)
\Gamma\left(  -1-\frac{t}{2}\right)  \left[  ik_{2}\cdot\partial X(1)\right]
^{1+\frac{t}{2}}e^{ikX(1)} \label{pomeron}%
\end{align}
where we have used%
\begin{equation}
\sum_{j=0}^{l}{\binom{l}{j}}\left(  \frac{2}{\tilde{t}}\right)  ^{j}\left(
-1-\frac{t}{2}\right)  _{j}=2^{l}(\tilde{t})^{-l}\ U\left(  -l,\frac{t}%
{2}+2-l,\frac{\tilde{t}}{2}\right)  . \label{equality}%
\end{equation}
One notes that the exponent of $\left[  ik_{2}\cdot\partial X(1)\right]
^{1+\frac{t}{2}}$ in Eq.(\ref{pomeron}) is mass level $N$ independent. This is
related to the fact that the well known $\sim s^{\alpha(t)}$ power-law
behavior of the four tachyon string scattering amplitude in the RR can be
extended to arbitrary higher string states and is mass level independent as
can be seen from Eq.(\ref{power}). This interesting result was first pointed
out in \cite{bosonic} and will be crucial to derive inter-mass level
recurrence relations among BPST vertex operators to be discussed later.

The BPST vertex operator in Eq.(\ref{pomeron}) leads to exactly the same
amplitude as in Eq.(\ref{amplitude-tensor}).

\section{Recurrence Relations}

For any confluent hypergeometric function $U(a,c,x)$ with parameters $(a,c)$
the four functions with parameters $(a-1,c),(a+1,c),(a,c-1)$ and $(a,c+1)$ are
called the contiguous functions. Recurrence relation exists between any such
function and any two of its contiguous functions. There are six recurrence
relations \cite{Slater}%
\begin{align}
U(a-1,c,x)-(2a-c+x)U(a,c,x)+a(1+a-c)U(a+1,c,x)  &  =0,\label{RR1}\\
(c-a-1)U(a,c-1,x)-(x+c-1))U(a,c,x)+xU(a,c+1,x)  &  =0,\label{RR2}\\
U(a,c,x)-aU(a+1,c,x)-U(a,c-1,x)  &  =0,\label{RR3}\\
(c-a)U(a,c,x)+U(a-1,c,x)-xU(a,c+1,x)  &  =0,\label{RR4}\\
(a+x)U(a,c,x)-xU(a,c+1,x)+a(c-a-1)U(a+1,c,x)  &  =0,\label{RR5}\\
(a+x-1)U(a,c,x)-U(a-1,c,x)+(1+a-c)U(a,c-1,x)  &  =0. \label{RR6}%
\end{align}
From any two of these six relations the remaining four recurrence relations
can be deduced.

The confluent hypergeometric function $U(a,c,x)$ with parameters $(a\pm m,c\pm
n)$ for $m,n=0,1,2...$are called associated functions. Again it can be shown
that there exist relations between any three associated functions, so that any
confluent hypergeometric function can be expressed in terms of any two of its
associated functions.

Recently it was shown \cite{LY} that Recurrence relations exist among higher
spin Regge string scattering amplitudes of different string states. The key to
derive these relations was to use recurrence relations and addition theorem of
Kummer functions. In view of the form of higher spin BPST vertex operators in
Eq.(\ref{pomeron}), one can easily calculate recurrence relations among higher
spin BPST vertex operators. By using the recurrence relation of Kummer
functions \cite{LY}, for example,
\begin{equation}
U\left(  -2,\frac{t}{2},\frac{t}{2}\right)  +\left(  \frac{t}{2}+1\right)
U(-1,\frac{t}{2},\frac{t}{2})-\frac{t}{2}U\left(  -1,\frac{t}{2}+1,\frac{t}%
{2}\right)  =0,
\end{equation}
one can obtain the following recurrence relation among BPST vertex operators
at mass level $M^{2}=2$
\begin{equation}
M\sqrt{-t}V_{BPST}^{(q_{1}=2)}-\frac{t}{2}V_{BPST}^{(p_{1}=1,q_{1}=1)}=0.
\label{Recurrence}%
\end{equation}
Rather than constant coefficients in the RR Regge stringy Ward identities
derived in \cite{LY}, the coefficients of this recurrence relation
Eq.(\ref{Recurrence}) among BPST vertex operators are kinematic variable
dependent, similar to BCJ relations among field theory amplitudes
\cite{BCJ1,BCJ2,BCJ3,BCJ4,BCJ5}. The recurrence relation among BPST vertex
operators in Eq.(\ref{Recurrence}) leads to the recurrence relation among
Regge string scattering amplitudes \cite{LY}
\begin{equation}
M\sqrt{-t}A^{(q_{1}=2)}-\frac{t}{2}A^{(p_{1}=1,q_{1}=1)}=0. \label{RRT1}%
\end{equation}

\section{More General Recurrence Relations}

To derive more general recurrence relations, we need to calculate BPST vertex
operators corresponding to the general higher spin states in Eq.(\ref{RR}). We
first calculate the BPST vertex operator correspong to the state%
\begin{equation}
\left\vert p_{n},r_{l}\right\rangle =\prod_{n=1}(\alpha_{-n}^{T})^{p_{n}}%
\prod_{m=1}(\alpha_{-l}^{L})^{r_{l}}|0\rangle. \label{state2}%
\end{equation}
The calculation is very similar to that of Eq.(\ref{state1}) up to some
modification. One can easily get that Eq.(\ref{dot}) is now replaced by
\begin{align}
V_{BPST}^{(p_{n};r_{l})}  &  =\left(  \frac{-\tilde{t}^{\prime}}{2M}\right)
^{r_{1}}\prod_{n=1}\left[  \sqrt{-t}(n-1)!\right]  ^{p_{n}}\prod_{l=2}\left[
\tilde{t}^{\prime}(l-1)!\left(  -\frac{1}{2M}\right)  \right]  ^{r_{l}%
}\nonumber\\
&  \cdot\sum_{j=0}^{r_{1}}{\binom{r_{1}}{j}}\left(  \frac{2iM\partial X\left(
1\right)  \cdot e^{L}}{\tilde{t}^{\prime}}\right)  ^{j}\Gamma\left(
-1-\frac{t}{2}+j\right)  \left[  ik_{2}\cdot\partial X(1)\right]  ^{1+\frac
{t}{2}-j}e^{ikX(1)}. \label{dot2}%
\end{align}
One notes that, in Eq.(\ref{dot2}), $M\partial X\left(  1\right)  \cdot
e^{L}\neq k_{2}\cdot\partial X(1)$ and, in contrast to Eq.(\ref{dot}), the two
factors with exponents $j$ and $-j$ do not cancel out. The BPST vertex
operator for this case thus reduces to
\begin{align}
V_{BPST}^{(p_{n};r_{l})}  &  =\left(  \frac{-1}{M}\right)  ^{r_{1}}\prod
_{n=1}\left[  \sqrt{-t}(n-1)!\right]  ^{p_{n}}\prod_{l=2}\left[  \tilde
{t}^{\prime}(l-1)!\left(  -\frac{1}{2M}\right)  \right]  ^{r_{l}}\nonumber\\
&  \cdot U\left(  -r_{1},\frac{t}{2}+2-r_{1},\frac{\tilde{t}^{\prime}}{2}%
\frac{e^{P}\cdot\partial X(1)}{e^{L}\cdot\partial X\left(  1\right)  }\right)
\Gamma\left(  -1-\frac{t}{2}\right)  \left[  ik_{2}\cdot\partial X(1)\right]
^{1+\frac{t}{2}}e^{ikX(1)}. \label{pomeron2}%
\end{align}
The BPST vertex operator in Eq.(\ref{pomeron2}) leads to the amplitude
\begin{align}
A^{(p_{n},r_{l})}  &  =\left(  -\frac{1}{M}\right)  ^{r_{1}}U\left(
-r_{1},\frac{t}{2}+2-r_{1},\frac{\tilde{t}^{\prime}}{2}\right)  \Gamma\left(
-1-\frac{t}{2}\right)  \left(  -\frac{s}{2}\right)  ^{1+\frac{t}{2}%
}\nonumber\\
&  \cdot\prod_{n=1}\left[  \sqrt{-t}(n-1)!\right]  ^{p_{n}}\prod_{l=2}\left[
\tilde{t}^{\prime}(l-1)!\left(  -\frac{1}{2M}\right)  \right]  ^{r_{l}},
\label{amplitude-tensor2}%
\end{align}
which is consistent with the one calculated in \cite{bosonic,RRsusy,LY}. Note
that the contribution of $\frac{e^{P}\cdot\partial X(1)}{e^{L}\cdot\partial
X\left(  1\right)  }$ in the correlation function reduces to $1$ in the Regge
limit by using first equations of Eq.(\ref{st}) and Eq.(\ref{st2}). One sees
that Eq.(\ref{amplitude-tensor2}) can be obtained from
Eq.(\ref{amplitude-tensor}) by doing the replacement $\tilde{t}\rightarrow$
$\tilde{t}^{\prime}$.

We are now ready to calculate the BPST vertex operator corresponding to the
most general Regge state in Eq.(\ref{RR}). Similar to the RR amplitude
calculated in \cite{LY}, the BPST vertex operator can be expressed in two
equivelent forms%
\begin{align}
V_{BPST}^{(p_{n};q_{m;}r_{l})}  &  =\prod_{n=1}\left[  \left(  n-1\right)
!\sqrt{-t}\right]  ^{p_{n}}\cdot\prod_{m=1}\left[  -\left(  m-1\right)
!\frac{\tilde{t}}{2M}\right]  ^{q_{m}}\cdot\prod_{l=2}\left[  \left(
l-1\right)  !\frac{\tilde{t}^{\prime}}{2M}\right]  ^{r_{l}}\nonumber\\
&  \quad\cdot\left(  \frac{1}{M}\right)  ^{r_{1}}\Gamma\left(  -1-\frac{t}%
{2}\right)  \left[  ik_{2}\cdot\partial X(1)\right]  ^{1+\frac{t}{2}%
}e^{ikX(1)}\nonumber\\
&  \cdot\sum_{i=0}^{q_{1}}\binom{q_{1}}{i}\left(  \frac{2}{\tilde{t}}\right)
^{i}\left(  -\frac{t}{2}-1\right)  _{i}U\left(  -r_{1},\frac{t}{2}%
+2-i-r_{1},\frac{\tilde{t}^{\prime}}{2}\frac{e^{P}\cdot\partial X(1)}%
{e^{L}\cdot\partial X\left(  1\right)  }\right) \label{factor1}\\
&  =\prod_{n=1}\left[  \left(  n-1\right)  !\sqrt{-t}\right]  ^{p_{n}}%
\cdot\prod_{m=2}\left[  -\left(  m-1\right)  !\frac{\tilde{t}}{2M}\right]
^{q_{m}}\cdot\prod_{l=1}\left[  \left(  l-1\right)  !\frac{\tilde{t}^{\prime}%
}{2M}\right]  ^{r_{l}}\nonumber\\
&  \cdot\left(  -\frac{1}{M}\right)  ^{q_{1}}\Gamma\left(  -1-\frac{t}%
{2}\right)  \left[  ik_{2}\cdot\partial X(1)\right]  ^{1+\frac{t}{2}%
}e^{ikX(1)}\nonumber\\
&  \cdot\sum_{j=0}^{r_{1}}\binom{r_{1}}{j}\left(  \frac{2}{\tilde{t}^{\prime}%
}\frac{e^{L}\cdot\partial X(1)}{e^{P}\cdot\partial X\left(  1\right)
}\right)  ^{j}\left(  -\frac{t}{2}-1\right)  _{j}U\left(  -q_{1},\frac{t}%
{2}+2-j-q_{1},\frac{\tilde{t}}{2}\right)  . \label{factor2}%
\end{align}
Either form Eq.(\ref{factor1}) or Eq.(\ref{factor2}) of the above BPST vertex
operator leads consistently to the amplitude calculated previously \cite{LY}%
\begin{align}
A^{(p_{n},q_{m}:r_{l})}  &  =\prod_{n=1}\left[  \left(  n-1\right)  !\sqrt
{-t}\right]  ^{p_{n}}\cdot\cdot\prod_{m=1}\left[  -\left(  m-1\right)
!\frac{\tilde{t}}{2M}\right]  ^{q_{m}}\cdot\prod_{l=2}\left[  \left(
l-1\right)  !\frac{\tilde{t}^{\prime}}{2M}\right]  ^{r_{l}}\nonumber\\
&  \cdot\left(  \frac{1}{M}\right)  ^{r_{1}}\Gamma\left(  -1-\frac{t}%
{2}\right)  \left(  -\frac{s}{2}\right)  ^{1+\frac{t}{2}}\\
&  \cdot\sum_{i=0}^{q_{1}}\binom{q_{1}}{i}\left(  \frac{2}{\tilde{t}}\right)
^{i}\left(  -\frac{t}{2}-1\right)  _{i}U\left(  -r_{1},\frac{t}{2}%
+2-i-r_{1},\frac{\tilde{t}^{\prime}}{2}\right) \label{RRamp1}\\
&  =\prod_{n=1}\left[  \left(  n-1\right)  !\sqrt{-t}\right]  ^{p_{n}}%
\cdot\prod_{m=2}\left[  -\left(  m-1\right)  !\frac{\tilde{t}}{2M}\right]
^{q_{m}}\cdot\prod_{l=1}\left[  \left(  l-1\right)  !\frac{\tilde{t}^{\prime}%
}{2M}\right]  ^{r_{l}}\nonumber\\
&  \cdot\left(  -\frac{1}{M}\right)  ^{q_{1}}\Gamma\left(  -1-\frac{t}%
{2}\right)  \left(  -\frac{s}{2}\right)  ^{1+\frac{t}{2}}\nonumber\\
&  \cdot\sum_{j=0}^{r_{1}}\binom{r_{1}}{j}\left(  \frac{2}{\tilde{t}^{\prime}%
}\right)  ^{j}\left(  -\frac{t}{2}-1\right)  _{j}U\left(  -q_{1},\frac{t}%
{2}+2-j-q_{1},\frac{\tilde{t}}{2}\right)  . \label{RRamp2}%
\end{align}

One can now derive more general recurrence relations among BPST vertex
operators. As an example, the three BPST vertex operators $V_{BPST}^{q_{1}=3}%
$, $V_{BPST}^{p_{1}=1,q_{1}=2}$ and $V_{BPST}^{q_{1}=2,r_{1}=1}$ can be
calculated by using Eq.(\ref{factor2}) to be%
\begin{align}
V_{BPST}^{(q_{1}=3)}  &  =\left(  -\frac{1}{M}\right)  ^{3}\Gamma\left(
-1-\frac{t}{2}\right)  \left[  ik_{2}\cdot\partial X(1)\right]  ^{1+\frac
{t}{2}}e^{ikX(1)}U\left(  -3,\frac{t}{2}-1,\frac{t}{2}-1\right)  ,\\
V_{BPST}^{(p_{1}=1,q_{1}=2)}  &  =\left(  -\frac{1}{M}\right)  ^{2}\sqrt
{-t}\Gamma\left(  -1-\frac{t}{2}\right)  \left[  ik_{2}\cdot\partial
X(1)\right]  ^{1+\frac{t}{2}}e^{ikX(1)}U\left(  -2,\frac{t}{2},\frac{t}%
{2}-1\right)  ,\\
V_{BPST}^{(q_{1}=2,r_{1}=1)}  &  =\frac{t+6}{2M}\left(  -\frac{1}{M}\right)
^{2}\Gamma\left(  -1-\frac{t}{2}\right)  \left[  ik_{2}\cdot\partial
X(1)\right]  ^{1+\frac{t}{2}}e^{ikX(1)}\nonumber\\
&  \left[  U\left(  -2,\frac{t}{2},\frac{t}{2}-1\right)  +\frac{2}{t+6}\left(
-\frac{t}{2}-1\right)  U\left(  -2,\frac{t}{2}-1,\frac{t}{2}-1\right)
\frac{e^{L}\cdot\partial X(1)}{e^{P}\cdot\partial X\left(  1\right)  }\right]
.
\end{align}
The recurrence relation among Kummer functions derived from Eq.(\ref{RR4})
\cite{LY}%
\begin{equation}
U\left(  -3,\frac{t}{2}-1,\frac{t}{2}-1\right)  +\left(  \frac{t}{2}+1\right)
U(-2,\frac{t}{2}-1,\frac{t}{2}-1)-\left(  \frac{t}{2}-1\right)  U\left(
-2,\frac{t}{2},\frac{t}{2}-1\right)  =0 \label{RRTT}%
\end{equation}
leads to the following recurrence relation among BPST vertex operators at mass
level $M^{2}=4$%
\begin{align}
M\sqrt{-t}e^{L}\cdot\partial X\left(  1\right)  V_{BPST}^{q_{1}=3}+M\sqrt
{-t}e^{P}\cdot\partial X(1)V_{BPST}^{q_{1}=2,r_{1}=1}  & \nonumber\\
-\left[  \left(  \frac{t}{2}+3\right)  e^{P}\cdot\partial X(1)-\left(
\frac{t}{2}-1\right)  e^{L}\cdot\partial X\left(  1\right)  \right]
V_{BPST}^{p_{1}=1,q_{1}=2}  &  =0. \label{RRT2}%
\end{align}
In addition to the $t$ dependence, the coefficients of the recurrence relation
in Eq.(\ref{RRT2}) are operator dependent. The recurrence relation among BPST
vertex operators in Eq.(\ref{RRT2}) leads to the recurrence relation among
Regge string scattering amplitudes \cite{LY}%
\begin{equation}
M\sqrt{-t}A^{(q_{1}=3)}-4A^{(p_{1}=1,q_{1}=2)}+M\sqrt{-t}A^{(q_{1}=2,r_{1}%
=1)}=0.
\end{equation}
\qquad

For the next example, we construct an inter-mass level recurrence relation for
BPST vertex operators at mass level $M^{2}=2,4.$ We begin with the addition
theorem of Kummer function \cite{Slater}%
\begin{equation}
U(a,c,x+y)=\sum_{k=0}^{\infty}\frac{1}{k!}\left(  a\right)  _{k}(-1)^{k}%
y^{k}U(a+k,c+k,x)
\end{equation}
which terminates to a finite sum for a nonpositive integer $a.$ By taking, for
example, $a=-1,c=\frac{t}{2}+1,x=\frac{t}{2}-1$ and $y=1,$ the theorem gives
\cite{LY}%
\begin{equation}
U\left(  -1,\frac{t}{2}+1,\frac{t}{2}\right)  -U\left(  -1,\frac{t}{2}%
+1,\frac{t}{2}-1\right)  -U\left(  0,\frac{t}{2}+2,\frac{t}{2}-1\right)  =0.
\label{inter}%
\end{equation}
Eq.(\ref{inter}) leads to an inter-mass level recurrence relation among BPST
vertex operators
\begin{equation}
M(2)(t+6)V_{BPST}^{(p_{1}=1,q_{1}=1)}-2M(4)^{2}\sqrt{-t}V_{BPST}%
^{(q_{1}=1,r_{2}=1)}+2M(4)V_{BPST}^{(p_{1}=1,r_{2}=1)}=0 \label{RRT3}%
\end{equation}
where \ masses $M(2)=\sqrt{2},M(4)=\sqrt{4}=2,$ and $V_{BPST}^{p_{1}%
=1,q_{1}=1}$ is a BPST vertex operator at mass level $M^{2}=2$, and
$V_{BPST}^{q_{1}=1,r_{2}=1}$, $V_{BPST}^{p_{1}=1,r_{2}=1}$are BPST vertex
operators at mass levels $M^{2}=4$ respectively. In deriving Eq.(\ref{RRT3}),
it is important to use the fact that the exponent of $\left[  ik_{2}%
\cdot\partial X(1)\right]  ^{1+\frac{t}{2}}$ in the BPST vertex operator in
Eq.(\ref{factor2}) is mass level $N$ independent as mentioned in the paragraph
after Eq.(\ref{equality}). The recurrence relation among BPST vertex operators
in Eq.(\ref{RRT3}) leads to the recurrence relation among Regge string
scattering amplitudes \cite{LY}%
\begin{equation}
M(2)(t+6)A^{(p_{1}=1,q_{1}=1)}-2M(4)^{2}\sqrt{-t}A^{(q_{1}=1,r_{2}%
=1)}+2M(4)A^{(p_{1}=1,r_{2}=1)}=0.
\end{equation}

In \cite{LY}, it was shown that, at each fixed mass level, each Kummer
function in the summation of Eq.(\ref{RRamp2}) can be expressed in terms of
Regge string scattering amplitudes $A^{(p_{n},q_{m}:r_{l})}$ at the same mass
level. Moreover, although for general values of $a$, the best one can obtain
from recurrence relations of Kummer function $U(a,c,x)$ is to express any
Kummer function in terms of any two of its associated function, for
nonpositive integer values of $a$ in the RR string amplitude case however,
$U(a,c,x)$ can be fixed up to an overall factor by using Kummer function
recurrence relations \cite{LY}. As a result, all Regge string scattering
amplitudes can be algebraically solved by Kummer function recurrence relations
up to multiplicative factors. An important application of the above properties
is the construction of an infinite number of recurrence relations among Regge
string scattering amplitudes. One can use the recurrence relations of Kummer
functions Eq.(\ref{RR1}) to Eq.(\ref{RR6}) to systematically construct
recurrence relations among Regge string scattering amplitudes.

In view of the form of BPST vertex operators calculated in Eq.(\ref{factor2}),
one can similarly solve \cite{LY} all Kummer functions $U(a,c,x)$ in
Eq.(\ref{factor2}) in terms of BPST vertex operators and use the recurrence
relations of Kummer functions Eq.(\ref{RR1}) to Eq.(\ref{RR6}) to
systematically construct an infinite number of recurrence relations among BPST
vertex operators. Moreover, the forms of all BPST vertex operators can be
fixed by these recurrence relations up to multiplicative factors. These
recurrence relations among BPST vertex operators are dual to linear relations
or symmetries among high-energy fixed angle string scattering amplitudes
discovered previously \cite{ChanLee1,ChanLee2,PRL,CHLTY}.

We illustrate the prescription here to construct other examples of recurrence
relations among BPST vertex operators at mass level $M^{2}=4.$ Generalization
to arbitrary mass levels will be given in the next section. There are $22$
BPST vertex operators for the mass level $M^{2}=4.$ We first consider the
group of BPST vertex operators with $q_{1}=0,$ $(V_{BPST}^{TTT},V_{BPST}%
^{LTT},V_{BPST}^{LLT},V_{BPST}^{LLL})$ \cite{LY}. The corresponding $r_{1}$
for each BPST vertex operator are $(0,1,2,3)$. Here we use a new notation for
BPST vertex operator, for example, $V_{BPST}^{LLT}\equiv V_{BPST}%
^{(p_{1}=1,r_{1}=2)}$ ,$V_{BPST}^{LT}=V_{BPST}^{(p_{1}=1,r_{2}=1)}$and
$V_{BPST}^{TL}=V_{BPST}^{(p_{2}=1,r_{1}=1)}$etc. By using Eq.(\ref{factor2}),
one can easily calculate that%
\begin{align}
V_{BPST}^{TTT}  &  =\left(  \sqrt{-t}\right)  ^{3}\Gamma\left(  -1-\frac{t}%
{2}\right)  \left[  ik_{2}\cdot\partial X(1)\right]  ^{1+\frac{t}{2}%
}e^{ikX(1)}U\left(  0,\frac{t}{2}+2,\frac{t}{2}-1\right)  ,\\
V_{BPST}^{LTT}  &  =\frac{t+6}{2M}\left(  \sqrt{-t}\right)  ^{2}\Gamma\left(
-1-\frac{t}{2}\right)  \left[  ik_{2}\cdot\partial X(1)\right]  ^{1+\frac
{t}{2}}e^{ikX(1)}\nonumber\\
&  \cdot\left[  U\left(  0,\frac{t}{2}+2,\frac{t}{2}-1\right)  +\frac{2}%
{t+6}\left(  -\frac{t}{2}-1\right)  U\left(  0,\frac{t}{2}+1,\frac{t}%
{2}-1\right)  \frac{e^{L}\cdot\partial X(1)}{e^{P}\cdot\partial X\left(
1\right)  }\right]  ,\\
V_{BPST}^{LLT}  &  =(\frac{t+6}{2M})^{2}\left(  \sqrt{-t}\right)
\Gamma\left(  -1-\frac{t}{2}\right)  \left[  ik_{2}\cdot\partial X(1)\right]
^{1+\frac{t}{2}}e^{ikX(1)}\nonumber\\
&  \cdot\left[
\begin{array}
[c]{c}%
U\left(  0,\frac{t}{2}+2,\frac{t}{2}-1\right)  +\frac{4}{t+6}\left(  -\frac
{t}{2}-1\right)  U\left(  0,\frac{t}{2}+1,\frac{t}{2}-1\right)  \frac
{e^{L}\cdot\partial X(1)}{e^{P}\cdot\partial X\left(  1\right)  }\\
+(\frac{2}{t+6})^{2}\left(  -\frac{t}{2}-1\right)  (-\frac{t}{2})U\left(
0,\frac{t}{2},\frac{t}{2}-1\right)  \left[  \frac{e^{L}\cdot\partial
X(1)}{e^{P}\cdot\partial X\left(  1\right)  }\right]  ^{2}%
\end{array}
\right]  ,\\
V_{BPST}^{LLL}  &  =(\frac{t+6}{2M})^{3}\Gamma\left(  -1-\frac{t}{2}\right)
\left[  ik_{2}\cdot\partial X(1)\right]  ^{1+\frac{t}{2}}e^{ikX(1)}\nonumber\\
&  \cdot\left[
\begin{array}
[c]{c}%
U\left(  0,\frac{t}{2}+2,\frac{t}{2}-1\right)  +\frac{6}{t+6}\left(  -\frac
{t}{2}-1\right)  U\left(  0,\frac{t}{2}+1,\frac{t}{2}-1\right)  \frac
{e^{L}\cdot\partial X(1)}{e^{P}\cdot\partial X\left(  1\right)  }\\
+3(\frac{2}{t+6})^{2}\left(  -\frac{t}{2}-1\right)  (-\frac{t}{2})U\left(
0,\frac{t}{2},\frac{t}{2}-1\right)  \left[  \frac{e^{L}\cdot\partial
X(1)}{e^{P}\cdot\partial X\left(  1\right)  }\right]  ^{2}\\
+(\frac{2}{t+6})^{3}\left(  -\frac{t}{2}-1\right)  (-\frac{t}{2})\left(
-\frac{t}{2}+1\right)  U\left(  0,\frac{t}{2}-1,\frac{t}{2}-1\right)  \left[
\frac{e^{L}\cdot\partial X(1)}{e^{P}\cdot\partial X\left(  1\right)  }\right]
^{3}%
\end{array}
\right]  .
\end{align}
From the above equations, one can easily see that $U\left(  0,\frac{t}%
{2}+2,\frac{t}{2}-1\right)  $ can be expressed in terms of $V_{BPST}^{TTT}$,
$U\left(  0,\frac{t}{2}+1,\frac{t}{2}-1\right)  $ can be expressed in terms of
$(V_{BPST}^{TTT},V_{BPST}^{LTT})$, $U\left(  0,\frac{t}{2},\frac{t}%
{2}-1\right)  $ can be expressed in terms of $(V_{BPST}^{TTT},V_{BPST}%
^{LTT},V_{BPST}^{LLT})$, and finally $U\left(  0,\frac{t}{2}-1,\frac{t}%
{2}-1\right)  $ can be expressed in terms of $(V_{BPST}^{TTT},V_{BPST}%
^{LTT},V_{BPST}^{LLT},V_{BPST}^{LLL})$. We have%
\begin{align}
U\left(  0,\frac{t}{2}+2,\frac{t}{2}-1\right)   &  =\Omega^{-1}\left(
\sqrt{-t}\right)  ^{-3}V_{BPST}^{TTT},\\
U\left(  0,\frac{t}{2}+1,\frac{t}{2}-1\right)   &  =\Omega^{-1}\left(
\sqrt{-t}\right)  ^{-3}\frac{t+6}{t+2}\left[  \frac{e^{P}\cdot\partial
X(1)}{e^{L}\cdot\partial X\left(  1\right)  }\right] \nonumber\\
&  \cdot\left[  V_{BPST}^{TTT}-\frac{2M}{t+6}\sqrt{-t}V_{BPST}^{LTT}\right]
,\\
U\left(  0,\frac{t}{2},\frac{t}{2}-1\right)   &  =\Omega^{-1}\left(  \sqrt
{-t}\right)  ^{-3}\frac{(t+6)^{2}}{t(t+2)}\left[  \frac{e^{P}\cdot\partial
X(1)}{e^{L}\cdot\partial X\left(  1\right)  }\right]  ^{2}\nonumber\\
&  \cdot\left[  V_{BPST}^{TTT}-2\frac{2M}{t+6}\sqrt{-t}V_{BPST}^{LTT}+\left(
\frac{2M}{t+6}\sqrt{-t}\right)  ^{2}V_{BPST}^{LLT}\right]  ,\\
U\left(  0,\frac{t}{2}-1,\frac{t}{2}-1\right)   &  =\Omega^{-1}\left(
\sqrt{-t}\right)  ^{-3}\frac{(t+6)^{3}}{t(t^{2}-4)}\left[  \frac{e^{P}%
\cdot\partial X(1)}{e^{L}\cdot\partial X\left(  1\right)  }\right]
^{3}\nonumber\\
&  \cdot\left[
\begin{array}
[c]{c}%
V_{BPST}^{TTT}-3\frac{2M}{t+6}\sqrt{-t}V_{BPST}^{LTT}\\
+3\left(  \frac{2M}{t+6}\sqrt{-t}\right)  ^{2}V_{BPST}^{LLT}-\left(  \frac
{2M}{t+6}\sqrt{-t}\right)  ^{3}V_{BPST}^{LLL}%
\end{array}
\right]  \label{higher}%
\end{align}
where $\Omega\equiv\Gamma\left(  -1-\frac{t}{2}\right)  \left[  ik_{2}%
\cdot\partial X(1)\right]  ^{1+\frac{t}{2}}e^{ikX(1)}$. To derive an example
of recurrence relation, one notes that Eq.(\ref{RR2}) gives%
\begin{equation}
\frac{t}{2}U\left(  0,\frac{t}{2},\frac{t}{2}-1\right)  -(t-1)U\left(
0,\frac{t}{2}+1,\frac{t}{2}-1\right)  +(\frac{t}{2}-1)U\left(  0,\frac{t}%
{2}+2,\frac{t}{2}-1\right)  =0,
\end{equation}
which leads to the recurrence relation among BPST vertex operators%
\begin{align}
\left[  \left(  \frac{t}{2}-1\right)  -\frac{(t-1)(t+6)}{t+2}\frac{e^{P}%
\cdot\partial X(1)}{e^{L}\cdot\partial X\left(  1\right)  }+\frac{(t+6)^{2}%
}{2(t+2)}\left[  \frac{e^{P}\cdot\partial X(1)}{e^{L}\cdot\partial X\left(
1\right)  }\right]  ^{2}\right]  V_{BPST}^{TTT}  & \nonumber\\
+\left[  \frac{(t-1)}{t+2}\frac{e^{P}\cdot\partial X(1)}{e^{L}\cdot\partial
X\left(  1\right)  }-\frac{(t+6)}{(t+2)}\left[  \frac{e^{P}\cdot\partial
X(1)}{e^{L}\cdot\partial X\left(  1\right)  }\right]  ^{2}\right]
(2M\sqrt{-t})V_{BPST}^{LTT}  & \nonumber\\
+\left[  \frac{1}{2(t+2)}\left[  \frac{e^{P}\cdot\partial X(1)}{e^{L}%
\cdot\partial X\left(  1\right)  }\right]  ^{2}\right]  (2M\sqrt{-t}%
)^{2}V_{BPST}^{LLT}  &  =0. \label{BPST}%
\end{align}
Again one can use Eq.(\ref{BPST}) to deduce recurrence relation among Regge
string scattering amplitudes%
\begin{equation}
(t+22)A^{(p_{1}=3)}-14M\sqrt{-t}A^{(p_{1}=2,r_{1}=1)}+2M^{2}(\sqrt{-t}%
)^{2}A^{(p_{1}=1,r_{1}=2)}=0. \label{LY}%
\end{equation}
Other recurrence relations of Kummer functions can be used to derive more
recurrence relations among BPST vertex operators. For example, Eq.(\ref{RR2})
gives a recurrence relation of $U\left(  0,\frac{t}{2}+1,\frac{t}{2}-1\right)
$ and its associated functions $U\left(  0,\frac{t}{2}-1,\frac{t}{2}-1\right)
$ and $U\left(  0,\frac{t}{2}+2,\frac{t}{2}-1\right)  $%
\begin{equation}
tU\left(  0,\frac{t}{2}-1,\frac{t}{2}-1\right)  -(3t-4)U\left(  0,\frac{t}%
{2}+1,\frac{t}{2}-1\right)  +2(t-2)U\left(  0,\frac{t}{2}+2,\frac{t}%
{2}-1\right)  =0,
\end{equation}
which leads to the recurrence relation among BPST vertex operators%
\begin{align}
\left[  2(t-2)-\frac{(3t-4)(t+6)}{t+2}\frac{e^{P}\cdot\partial X(1)}%
{e^{L}\cdot\partial X\left(  1\right)  }+\frac{(t+6)^{3}}{(t^{2}-4)}\left[
\frac{e^{P}\cdot\partial X(1)}{e^{L}\cdot\partial X\left(  1\right)  }\right]
^{3}\right]  V_{BPST}^{TTT}  & \nonumber\\
+\left[  \frac{(3t-4)}{t+2}\frac{e^{P}\cdot\partial X(1)}{e^{L}\cdot\partial
X\left(  1\right)  }-3\frac{(t+6)^{2}}{(t^{2}-4)}\left[  \frac{e^{P}%
\cdot\partial X(1)}{e^{L}\cdot\partial X\left(  1\right)  }\right]
^{3}\right]  (2M\sqrt{-t})V_{BPST}^{LTT}  & \nonumber\\
+\left[  \frac{3(t+6)}{(t^{2}-4)}\left[  \frac{e^{P}\cdot\partial X(1)}%
{e^{L}\cdot\partial X\left(  1\right)  }\right]  ^{3}\right]  (2M\sqrt
{-t})^{2}V_{BPST}^{LLT}  & \nonumber\\
-\left[  \frac{1}{(t^{2}-4)}\left[  \frac{e^{P}\cdot\partial X(1)}{e^{L}%
\cdot\partial X\left(  1\right)  }\right]  ^{3}\right]  (2M\sqrt{-t}%
)^{3}V_{BPST}^{LLL}  &  =0. \label{BPST2}%
\end{align}
one can use Eq.(\ref{BPST2}) to deduce recurrence relation among Regge string
scattering amplitudes%
\begin{align}
(3t^{2}+76t+92)A^{(p_{1}=3)}-2(23t+50)M\sqrt{-t}A^{(p_{1}=2,r_{1}=1)}  &
\nonumber\\
+6M^{2}(t+6)(\sqrt{-t})^{2}A^{(p_{1}=1,r_{1}=2)}-4M^{3}(\sqrt{-t}%
)^{3}A^{(r_{1}=3)}  &  =0.
\end{align}
Similarly, we can consider groups of BPST vertex operators $(V_{BPST}%
^{PT},V_{BPST}^{PL})$, $(V_{BPST}^{LT},V_{BPST}^{LL})$ and $(V_{BPST}%
^{TT},V_{BPST}^{TL})$ with $q_{1}=0$; group of BPST vertex operators
$(V_{BPST}^{PTT},V_{BPST}^{PLT},V_{BPST}^{PLL})$ with $q_{1}=1$ and group of
BPST vertex operators $(V_{BPST}^{PPT},V_{BPST}^{PPL})$ with $q_{1}=2$. All
the remaining $7$ BPST vertex operators are with $r_{1}=0$, and each BPST
vertex operators contains only one Kummer function. Thus all Kummer functions
involved at mass level $M^{2}=4$ can be algebraically solved and expressed in
terms of BPST vertex operators. One can then use recurrence relations of
Kummer functions to derive more recurrence relations among BPST vertex operators.

\section{Arbitrary Mass Levels}

In this section, we solve the Kummer functions in terms of the highest spin
string states scattering amplitudes for arbitrary mass levels. The highest
spin string states\ at the mass level $M^{2}=2\left(  N-1\right)  $ are
defined as%
\begin{equation}
\left\vert N-q_{1}-r_{1},q_{1},r_{1}\right\rangle =\left(  \alpha_{-1}%
^{T}\right)  ^{N-q_{1}-r_{1}}\left(  \alpha_{-1}^{P}\right)  ^{q_{1}}\left(
\alpha_{-1}^{L}\right)  ^{r_{1}}|0,k\rangle
\end{equation}
where only $\alpha_{-1}$ operator appears. The highest spin string states BPST
vertex operators can be easily obtained from Eq.(\ref{factor2}) as%
\begin{align}
&  \left(  V^{T}\right)  ^{N-q_{1}-r_{1}}\left(  V^{P}\right)  ^{q_{1}}\left(
V^{L}\right)  ^{r_{1}}\equiv V_{BPST}^{(N-q_{1}-r_{1},q_{1},r_{1})}\nonumber\\
&  =\Gamma\left(  -\frac{t}{2}-1\right)  \left[  ik_{2}\cdot\partial
X(1)\right]  ^{1+\frac{t}{2}}e^{ikX(1)}\left(  \sqrt{-t}\right)
^{N-q_{1}-r_{1}}\left(  -\frac{1}{M}\right)  ^{q_{1}}\left(  \frac{\tilde
{t}^{\prime}}{2M}\right)  ^{r_{1}}\nonumber\\
&  \cdot\sum_{j=0}^{r_{1}}\binom{r_{1}}{j}\left(  \frac{2}{\tilde{t}^{\prime}%
}\frac{e^{L}\cdot\partial X(1)}{e^{P}\cdot\partial X\left(  1\right)
}\right)  ^{j}\left(  -\frac{t}{2}-1\right)  _{j}U\left(  -q_{1},\frac{t}%
{2}+2-j-q_{1},\frac{\tilde{t}}{2}\right)  . \label{highest spin vertex}%
\end{align}
In view of the form of Eq.(\ref{higher}), we can solve the Kummer function
from Eq.(\ref{highest spin vertex}) and express it in terms of the highest
spin BPST vertex operators as%
\begin{align}
U\left(  -q_{1},\frac{t}{2}+2-q_{1}-r_{1},\frac{\tilde{t}}{2}\right)   &
=\frac{\Gamma\left(  -\frac{t}{2}-1\right)  }{\left(  -\frac{t}{2}-1\right)
_{r_{1}}}\left[  ik_{2}\cdot\partial X(1)\right]  ^{1+\frac{t}{2}}%
e^{ikX(1)}\nonumber\\
&  \cdot\left(  -MV^{P}\right)  ^{q_{1}}\left(  \frac{V^{T}}{\sqrt{-t}%
}\right)  ^{N-q_{1}}\left[  \frac{e^{P}\cdot\partial X(1)}{e^{L}\cdot\partial
X\left(  1\right)  }\left(  \sqrt{-t}M\frac{V^{L}}{V^{T}}-\frac{\tilde
{t}^{\prime}}{2}\right)  \right]  ^{r_{1}}. \label{Kummer functions}%
\end{align}
Putting the Kummer functions (\ref{Kummer functions}) into the recurrence
relations (\ref{RR1}-\ref{RR6}), we can then obtain recurrence relations among
BPST vertex operators.

Let us consider, for example, the recurrence relation%
\begin{equation}
\left(  c-a-1\right)  U(a,c-1,x)-(x+c-1))U(a,c,x)+xU(a,c+1,x)=0.
\end{equation}
With%
\begin{equation}
a=-q_{1},c=\frac{t}{2}+1-q_{1}-r_{1},x=\frac{\tilde{t}}{2}=\frac{t-M^{2}+2}%
{2},
\end{equation}
the above recurrence relation becomes%
\begin{align}
\left(  \frac{t}{2}-r_{1}\right)  U\left(  -q_{1},\frac{t}{2}-q_{1}%
-r_{1},\frac{\tilde{t}}{2}\right)   & \nonumber\\
-\left(  \frac{\tilde{t}}{2}+\frac{t}{2}-q_{1}-r_{1}\right)  U\left(
-q_{1},\frac{t}{2}+1-q_{1}-r_{1},\frac{\tilde{t}}{2}\right)   & \nonumber\\
+\frac{\tilde{t}}{2}U\left(  -q_{1},\frac{t}{2}+2-q_{1}-r_{1},\frac{\tilde{t}%
}{2}\right)   &  =0.
\end{align}
Plug the Kummer functions (\ref{Kummer functions}) into the above recurrence
relation, we obtain the recurrence relation among BPST vertex operators at
general mass level $N$%
\begin{equation}
\left(  V^{P}\right)  ^{q_{1}}\left(  V^{T}\right)  ^{N-q_{1}}\left(
X\right)  ^{r_{1}}\left[  X^{2}+\left(  \frac{\tilde{t}}{2}+\frac{t}{2}%
-q_{1}-r_{1}\right)  X+\frac{\tilde{t}}{2}\left(  \frac{t}{2}+1-r_{1}\right)
\right]  =0
\end{equation}
where we have defined%
\begin{equation}
X\equiv\frac{e^{P}\cdot\partial X(1)}{e^{L}\cdot\partial X\left(  1\right)
}\left(  \sqrt{-t}M\frac{V^{L}}{V^{T}}-\frac{\tilde{t}^{\prime}}{2}\right)
=\frac{e^{P}\cdot\partial X(1)}{e^{L}\cdot\partial X\left(  1\right)  }\left(
\sqrt{-t}M\frac{V^{L}}{V^{T}}-\frac{t+M^{2}+2}{2}\right)  .
\end{equation}
As an example, at the mass level $M^{2}=4$ with $q_{1}=r_{1}=0$, we get%
\begin{equation}
\left(  V^{T}\right)  ^{3}\left[  X^{2}+\left(  t-1\right)  X+\left(
\frac{t^{2}}{4}-1\right)  \right]  =0 \label{example}%
\end{equation}
where%
\begin{equation}
X=\frac{e^{P}\cdot\partial X(1)}{e^{L}\cdot\partial X\left(  1\right)
}\left(  \sqrt{-t}M\frac{V^{L}}{V^{T}}-\frac{t+6}{2}\right)  .
\end{equation}
A simple calculation shows that Eq.(\ref{example}) is exactly the same as
Eq.(\ref{BPST}), and the same recurrence relation among Regge string
scattering amplitudes (\ref{LY}) follows.

\section{Discussion}

Although we focus here on the spin-dependence of the 4-point open-string
amplitudes, it is useful to briefly recall the generality of the BPST vertex
operator, which emphasizes on Regge factorization and can be applied to
arbitrary n-point amplitudes, $n\geq4$. A Regge limit is defined by singling
out a longitudinal direction, e.g., the $z$-axis, along which all momenta are
large while keeping transverse components, $p_{\perp}$, fixed. We separate
particles into two groups, the right-moving and left-moving, with large
$p_{+}$ and $p_{-}$ large respectively. Each can have $n_{R}$ and $n_{L}$
states, with $n_{R}+n_{L}=n$ and $n_{R},n_{L}\geq2$. Within each group,
relative momenta remain finite in the Regge limit. Any n-point open-string
amplitude can formally be expressed in a factorable form $A_{L,R}=\int
dw\langle W_{R}w^{L_{0}-2}W_{L}\rangle$, where $W_{R}$ and $W_{L}$ are
products of respective right-moving and left-moving vertex operators, with all
world-sheet integrations done except one, i.e., w. The last remaining
integration is such that the factor $w^{L_{0}}$ corresponds to overall
rescaling in world-sheet coordinates in $W_{L}.$ (For more details, see
\cite{BPST}.) In the Regge limit, the amplitude $A_{L,R}$ takes on a simply
factorized form and it can be expressed in terms of the BPST vertex operator
\begin{align}
A_{L,R}  &  =\langle W_{R}V^{-}\rangle\;\Pi(t)\;\langle V^{+}W_{L}%
\rangle\nonumber\\
&  =\langle W_{R,0}V^{-}\rangle\;\Big\{\Pi(t)s^{\alpha(t)}\Big\}\;\langle
V^{+}W_{L,0}\rangle
\end{align}
where $\alpha(t)$ is the leading Regge trajectory, with $\alpha^{\prime}=1/2$,
and $\Pi(t)$ is a Regge propagator, given by a Gamma function. Here $V^{\pm}$
are BPST vertex operators, which are "on-shell" along the leading trajectory.
This is the most general form of Regge factorization for any number of
external particles. The factors $\langle W_{R,0}V^{-}\rangle$ and $\langle
V^{+}W_{L,0}\rangle$ are generalized $(n_{R}+1)$- and $(n_{L}+1)$-point
on-shell amplitudes, evaluated in the respective rest-frame, with one external
line being on the leading Regge trajectory. Each, due to Mobius invariance,
involves $n_{R}-2$ and $n_{L}-2$ world sheet integrations.

We have studied in this paper the Regge behavior of four-point open-string
scattering amplitudes, with one particle having arbitrary high spin and three
other being tachyons, using the technique of BPST vertex operator. Since we
only work with 4-point amplitudes in this paper, $n_{r}=n_{l}=2,$ there is no
integration involved for $\langle W_{R,0} V^{-}\rangle$ and $\langle V^{+}
W_{L,0}\rangle$, due to Mobius invariance. In particular, $W_{L}$ involves two
tachyons. Since one can show that $\langle V^{+} W_{L,0}\rangle$ is simply a
constant, therefore, what we have calculated is simply $\langle W_{R,0}
V^{-}\rangle$. With $W_{R}$ a product of two vertex operators, one for a
tachyon and another for a string state with arbitrary spin. For brevity, we
have collectively referred to $W_{R,0} V^{-}$ as BPST vertex operators.
Generalization of our analysis to amplitudes for $n=5,6\cdots$ will be treated elsewhere.

We have derived in this paper an infinite number of recurrence relations among
these matrix elements of the BPST vertex operator between different string
states with different spins, which can be expressed in terms of Kummer
function of the second kind. These recurrence relations lead to the same
recurrence relations among Regge string scattering amplitudes recently
discovered in \cite{LY} by a more traditional method. We show that all Kummer
functions involved at each fixed mass level can be algebraically solved and
expressed in terms of BPST vertex operators. We give a prescription to
construct recurrence relations among BPST vertex operators. For illustration,
we calculate some examples of recurrence relations among BPST vertex operators
of different string states based on recurrence relations and addition theorem
of Kummer functions. We stress that although the higher spin BPST vertex
operators were considered in \cite{BPST,CCW}, the key observation on the
energy orders in the Regge limit from polarizations of higher spin states was
not discussed in \cite{BPST,CCW}. One can not obtain recurrence relations
among higher spin BPST vertex operators in the Regge limit without including
the energy orders from these higher spin polarizations.

The recurrence relations among BPST vertex operators lead to the recurrence
relations among Regge string scattering amplitudes. They are thus both closely
related to Regge stringy Ward identities \cite{LY} derived from decoupling of
Regge ZNS in the string spectrum. These recurrence relations are dual to
linear relations derived from ZNS or symmetries among high-energy fixed angle
string scattering amplitudes \cite{ChanLee1,ChanLee2,PRL,CHLTY}.

\section{\bigskip Acknowledgments}

We thank Marko Djuric, Song He, Yu-Ting Huang and Yoshihiro Mitsuka for
helpful discussions. This work is supported in part by the National Science
Council, 50 billions project of Ministry of Education, National Center for
Theoretical Sciences and S.T. Yau center of NCTU, Taiwan.

\end{document}